\documentclass[aps,prb,superscriptaddress,twocolumn,showpacs,tightenlines,
amsmath,amsfonts] {revtex4}

\usepackage{graphics,bm}

\begin{document}

\author{O. A. Tretiakov} 
\thanks{Present address: Department of Physics and Astronomy, The Johns
Hopkins University, Baltimore, Maryland 21218, USA}
\affiliation{Department of Physics, Duke University, Durham, North
Carolina 27708, USA}

\author{K. A. Matveev} 
\affiliation{Materials Science Division,
Argonne National Laboratory, Argonne, Illinois 60439, USA}
\affiliation{Department of Physics, Duke University, Durham, North
Carolina 27708, USA}

\date{January 10, 2006}

\title{Decay of metastable current states in one-dimensional resonant
  tunneling devices}

\begin{abstract}
  Current switching in a double-barrier resonant tunneling structure is
  studied in the regime where the current-voltage characteristic exhibits
  intrinsic bistability, so that in a certain range of bias two different
  steady states of current are possible.  Near the upper boundary $V_{th}$
  of the bistable region the upper current state is metastable, and
  because of the shot noise it eventually decays to the stable lower
  current state.  We find the time of this switching process in
  strip-shaped devices, with the width small compared to the length.  As
  the bias $V$ is tuned away from the boundary value $V_{th}$ of the
  bistable region, the mean switching time $\tau$ increases exponentially.
  We show that in long strips $\ln\tau \propto (V_{th} -V)^{5/4}$, whereas
  in short strips $\ln\tau \propto (V_{th} -V)^{3/2}$.  The
  one-dimensional geometry of the problem enables us to obtain
  analytically exact expressions for both the exponential and the
  prefactor of $\tau$.  Furthermore, we show that, depending on the
  parameters of the system, the switching can be initiated either inside
  the strip, or at its ends.
\end{abstract}

\pacs{73.40.Gk, 73.21.Ac, 73.50.Td}

\maketitle

\hyphenation{pre-factor}

\section{Introduction}

\begin{figure}[b]
  \resizebox{.45\textwidth}{!}{\includegraphics{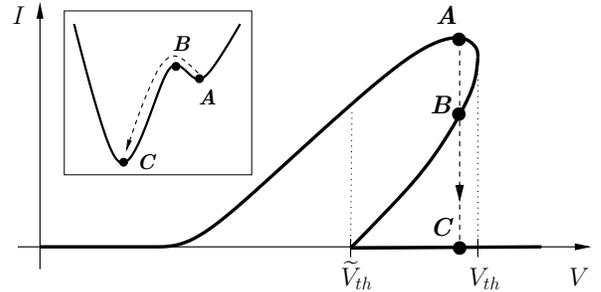}}
\caption{\label{fig:IVcurve} Current-voltage characteristic of the DBRTS.  
  The bistable region is present in the range of bias between ${\widetilde
    V}_{th}$ and $V_{th}$.  The inset shows a sketch of generic potential
  $u(n)$ in the bistable region.  The points $\textbf{\textit{A}}$ and
  $\textbf{\textit{C}}$ on the $I$-$V$ curve correspond to the local and
  global minima of $u(n)$, respectively.  The point $\textbf{\textit{B}}$
  on the unstable current branch corresponds to the maximum of $u$. }
\end{figure}

In the last two decades with advances of miniaturization techniques
various resonant tunneling structures became the subject of intensive
research.  It was experimentally observed that the current-voltage
characteristics of resonant tunneling devices, such as double-barrier
resonant tunneling structures~\cite{Goldman1, Alves1, Goldman2,
  Hayden:exp, Mendez} (DBRTS) and superlattices,~\cite{Grahn94, Grahn96,
  Grahn98, Teitsworth} exhibit intrinsic bistabilities.  Namely, it was
shown that for each value of bias in the bistable region of the $I$-$V$
curve the current can take two different values.  A bistable $I$-$V$ curve
of a double-barrier structure derived theoretically in
Ref.~\onlinecite{Blanter99} is depicted schematically in
Fig.~\ref{fig:IVcurve}.  It is theoretically established~\cite{paper1,
  dbrts2D} that near the boundary $V_{th}$ of the bistable region (e.g.,
at point $\textbf{\textit{A}}$) the upper current state is metastable.  In
recent experiments~\cite{Grahn98, Teitsworth} the switching from the
metastable to stable current state was studied in superlattices.  In
particular, the mean switching time $\tau$ was measured.

The problems of decay of metastable states were studied theoretically in
various fields, such as condensed matter physics,~\cite{Langer1,
  LangerAmb, MHalperin, Kurkijarvi, IordanskiiFinkel1, Jordan, Victora}
quantum field theory,~\cite{Voloshin, Coleman2, Coleman3, Kiselev,
  Selivanov} and chemical kinetics.~\cite{Dykman1, Kamenev} In the context
of resonant tunneling structures it was addressed in
Refs.~\onlinecite{paper1, dbrts2D}.  A typical double-barrier resonant
tunneling structure consists of three semiconducting layers of GaAs
separated by two insulating layers of GaAlAs.  In the narrow middle layer
of GaAs the electron motion in the direction normal to the layers is
quantized, so that a quantum well is formed.  In the bistable region
(Fig.~\ref{fig:IVcurve}) the two current states correspond to two
different values of electron density $n$ in the well.  The density $n$
exhibits shot noise fluctuations caused by random events of tunneling of
electrons in and out of the well.  Thus it becomes possible for the device
to switch from one current state to the other.  In
Refs.~\onlinecite{paper1, dbrts2D} it was shown that there are two regimes
of current switching.  In the case of relatively small samples the
electrons spread uniformly over the well due to diffusion.  Then the
switching occurs simultaneously in the entire area of the sample, and it
was found that the switching time $\tau$ is exponentially large, with
$\ln\tau \propto (V_{th}-V)^{3/2}$.  In larger samples the density in the
well is not uniform, and the switching occurs through nucleation
mechanism: it initiates in a small region of the quantum well of the
characteristic size $r_0$, which then spreads rapidly to the entire
sample.  The size of the critical nucleus of the stable current state was
found to be bias dependent, $r_0 \propto (V_{th}-V)^{-1/4}$.  It has been
shown that in the large-sample regime $\ln\tau \propto (V_{th}-V)$.

These results were obtained for samples whose two lateral dimensions are
comparable to each other.  In this paper we study the interesting case of
the devices of strip geometry with $w\ll L$, where $w$ is the width and
$L$ is the length of the strip.  In these devices there is a special
regime where the applied bias is such that $w\ll r_0 \ll L$.  Since the
scale $r_0$ gives the characteristic size of the density fluctuations, in
this case the density does not vary across the strip, but only along it.
Confining the switching process to one dimension alters its properties
significantly.  Similarly to the two-dimensional case, the switching time
grows exponentially when bias is tuned inside the bistable region.
However, the exponent follows a different dependence which is found in
Sec.~\ref{sec:1D}.

We show that in long strips the nucleation can occur either inside or at
the ends of the device.  It turns out that both nucleation regimes can be
observed, but the respective decay times are dramatically different.  To
compare them, one needs to calculate the switching times with the
prefactors.  Unlike the case of two-dimensional structures, in one
dimension the prefactors can be found exactly in the limit of long and
narrow strip, Sec.~\ref{end}.

The decay of the metastable current state is governed by shot noise,
and therefore switching is a stochastic process.  We describe this
process using the Fokker-Planck equation approach.  In
Sec.~\ref{sec:FPE} we introduce the Fokker-Planck equation for
tunneling and in-plane diffusion in DBRTS.  This equation is used to
calculate the prefactor of the mean switching time in strip-shaped
(Secs.~\ref{sec:end} and~\ref{inside1D}) and ring-shaped devices
(Sec.~\ref{ring}).  We summarize our results and discuss their
experimental implications in Sec.~\ref{discussion1D}.

\section {\label{sec:2} Stationary distribution function of electron density}

It was shown in Ref.~\onlinecite{dbrts2D} that the stationary distribution
function of electron density $n(\mathbf{r})$ in the quantum well takes the
form
\begin{equation}
\label{P0general}
P_{0}\{n\} = e^{-F\{n\}},\quad
F\{n\} = \int\!\! d^2 \mathbf{r} \left[ u(n) +\eta (\nabla n)^{2} \right].
\end{equation}
Here the integration is over the cross-section of the well.  In the case
of uniform density $F\{n\}$ is determined by the effective potential
$u(n)$, which describes the tunneling between the well and the leads.  The
gradient term in the functional $F\{n\}$ accounts for the diffusion in the
quantum well.  The constant $\eta$ is positive and proportional to the
in-plane conductivity,~\cite{dbrts2D} $\eta\propto\sigma$.  By suppressing
gradients of $n$, this term favors the states with uniform density.

The shape of the effective potential $u(n)$ in the bistable region is
schematically shown in the inset of Fig.~\ref{fig:IVcurve}.  It has the
local and global minima at points $\textbf{\textit{A}}$ and
$\textbf{\textit{C}}$, respectively.  In the case of uniform density $n$,
these minima result in peaks of the distribution function $P_{0}\{n\}$,
which correspond to the upper and lower branches of the $I$-$V$ curve.
Point $\textbf{\textit{B}}$ on the unstable current branch corresponds to
a maximum of $u$.

In the vicinity of the threshold voltage $V_{th}$ the effective
potential $u(n)$ can be approximated by a cubic polynomial
\begin{equation}
\label{u}
u(n) = - \alpha(n-n_{th}) + \frac{\gamma}{3} (n-n_{th})^{3},\quad
\alpha = a(V_{th}-V).
\end{equation}
Here $\gamma$ and $a$ are positive constants.  The voltage dependence of
$\alpha$ ensures that the local minimum of $u$, which corresponds to the
metastable state, disappears at the bistability threshold.  The threshold
density $n_{th}$ is defined as the density at point $V=V_{th}$, where the
local minimum $\textbf{\textit{A}}$ of $u(n)$ disappears by merging with
the maximum $\textbf{\textit{B}}$.

If the system is in the local minimum of $u(n)$, it will eventually decay
to the global minimum.  In the limit of large conductivity $\sigma$ or
small sample size the electron density $n$ in the quantum well is uniform,
and the gradient term in $F\{n\}$, Eq.~(\ref{P0general}), can be omitted.
In this case Eq.~(\ref{P0general}) is simplified: the distribution
function $P_0$ is described by the only variable $n$ and takes the form
$P_{0}(n) = e^{-Su(n)}$, where $S$ is the area of the sample.  Initially,
the system is in the local minimum of $u(n)$, see point
$\textbf{\textit{A}}$ in the inset of Fig.~\ref{fig:IVcurve}.  In order to
switch to the global minimum the system has to pass through point
$\textbf{\textit{B}}$.  As it follows from the expression for $P_0 (n)$,
the probability of reaching point $\textbf{\textit{B}}$ is exponentially
small, with the exponent determined by the barrier height
$S(u_{\textbf{\textit{B}}} -u_{\textbf{\textit{A}}})$.  The latter can be
easily found from the expansion~(\ref{u}), and the mean switching time
$\tau_0$ takes the form~\cite{dbrts2D}
\begin{equation}
\label{Tau0}
\tau_0 = \tau_0^* \exp\left(
\frac43\frac{Lw \alpha^{3/2}}{\gamma^{1/2}}\right).
\end{equation}
Here $\tau_0^*$ is a preexponential factor, and we assumed that the
cross-section of the sample has rectangular shape with the width $w$ and
the length $L$.

Expression~(\ref{Tau0}) is valid as long as $L,w \ll r_0$, where
\begin{equation}
\label{r0}
r_0 = \left( \frac{\eta^2}{\alpha\gamma} \right)^{1/4}
\propto (V_{th}-V)^{-1/4}
\end{equation}
has the meaning of the characteristic spatial scale of stochastic
fluctuations of electron density.~\cite{dbrts2D} The scale $r_0$ can
be tuned by changing the bias $V$.  If $L,w\gg r_0$ the switching
occurs according to the nucleation scenario.  In this case the
critical switching density is first achieved in a small part of the
sample of size $\sim r_0$.  After stochastic creation of the critical
nucleus, it grows rapidly in size until it occupies the entire sample.

In this paper we consider the case of a very narrow strip, $w\ll L$.
In the regime when the bias is such that $w\ll r_0$, the density may
change only along the strip, and the problem becomes
one-dimensional.\cite{footnote0}

In the following it will be convenient to express the density $n(x)$
in terms of a dimensionless function $z(\xi)$ that vanishes at the
minimum of $u(n)$, namely,
\begin{equation}
\label{nz}
n(x) = n_{\rm{min}}-2\sqrt{\frac{\alpha}{\gamma} }\, z(x/r_{0}).  
\end{equation}
Here the density at the minimum $n_{\rm{min}} = n_{th} + \sqrt{\alpha
  /\gamma}$.  Substituting Eq.~(\ref{nz}) into the functional $F$ in
  Eq.~(\ref{P0general}), we find
\begin{equation}
\label{P0}
F = U_{1}\int\! d\xi
\left(\frac{z'^{2}}{2} + \frac{z^{2}}{2} 
- \frac{z^{3}}{3}\right),
\end{equation}
where prime denotes differentiation with respect to the dimensionless
coordinate $\xi = x/r_0$ along the strip.  The characteristic value of the
functional $F$ is given by the parameter
\begin{equation}
  \label{U1}
  U_1 = \frac{8w\sqrt{\eta}\alpha^{5/4}}{\gamma^{3/4}}.
\end{equation}
Its value depends on bias as $U_1\propto (V_{th}-V)^{5/4}$.

\section {\label{sec:1D} The exponent of the mean switching time in long 
strips}

In this section we consider the regime $r_0 \lesssim L$.  In this case,
the density fluctuations along the strip result in a significant change of
the mean switching time, and the result~(\ref{Tau0}) is no longer
applicable.  The one-dimensional nature of this problem allows us to
obtain an explicit expression for the exponent of the mean switching time
$\tau$.

We begin by finding the minimum and the saddle points of the functional
$F$ in Eq.~(\ref{P0}).  They can be found from the condition $\delta
F/\delta z = 0$, that is,
\begin{equation}
\label{equation_1D}
-\frac{d^{2} z}{d \xi^{2}} +z -z^{2} = 0.
\end{equation}
This equation should be solved with the boundary conditions $z'(0) =
z'(\Lambda) = 0$ which account for the fact that there is no current
flowing through the ends of the strip.  Here we have introduced the
dimensionless length of the strip $\Lambda = L/r_0$.

\begin{figure}
\resizebox{.48\textwidth}{!}{\includegraphics{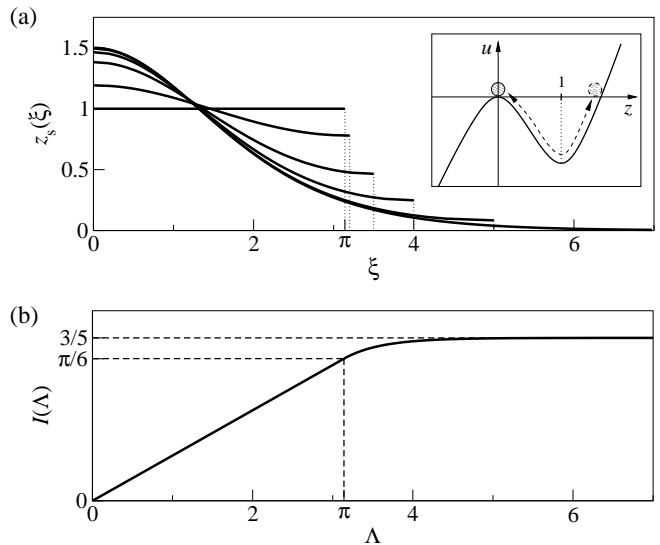}}
\\[2ex]
\resizebox{.48\textwidth}{!}{\includegraphics{exponent.eps}}
\caption{\label{fig:saddle1D}(a) Saddle point solution $z_s (\xi)$ for
  strips of length $\Lambda = \pi ,3.2, 3.5, 4, 5,$ and $\infty$. The
  allowed motion of a classical particle in the potential $u(z) = -z^2 /2
  +z^3 /3$ is shown in the inset. (b) The dependence $I(\Lambda)$ defined
  by Eq.~(\ref{Fsaddle}).  At $\Lambda < \pi$ it is linear: $I(\Lambda) =
  \Lambda/6$; while for $\Lambda > \pi$ it is described by
  Eq.~(\ref{intI}).  At $\Lambda\to\infty$ the saddle point $z_s (\xi)$ is
  given by Eq.~(\ref{saddle_1Dinf}), and $I = 3/5$.}
\end{figure}

Equation~(\ref{equation_1D}) can be interpreted as the equation of motion
of a classical particle with unit mass in the potential $u(z) = -z^2 /2
+z^3 /3$, see the inset of Fig.~\ref{fig:saddle1D}(a).  In this analogy
the coordinate $\xi$ plays the role of time.  There are two obvious
solutions, $z(\xi) = 0$ and $z(\xi) =1$, corresponding to the particle
staying at the maximum and minimum of $u(z)$, respectively.  The minimum
of $F\{z\}$ is obviously given by $z (\xi) = 0$, since we defined $z$ in
such a way that $z=0$ at the minimum.  The other solution, $z(\xi)= 1$, is
a saddle point of $F\{z\}$.

Apart from the two trivial solutions, equation~(\ref{equation_1D}) may
have $\xi$-dependent solutions corresponding to a moving particle.  The
boundary conditions $z'(0) = z'(\Lambda) = 0$ require zero velocity at the
moments $\xi =0$ and $\xi =\Lambda$.  Thus the particle performs
oscillatory motion between turning points $z(0)$ and $z(\Lambda)$.  If the
particle starts at $z(0)=c_0$ in the range $0\le c_0 \le 3/2$, one can
easily find the other turning point $c$ from the condition $u(c) = u(c_0
)$.  This equation has two solutions:
\begin{equation*}
c_{\pm} = \frac{3 - 2c_0 \pm \sqrt{-12c_0^2 +12c_0 +9}}{4}.
\end{equation*}
The turning point $z(\Lambda)$ corresponds to the positive root $c_+$.

The time $\Lambda$ required for the particle to travel from one turning
point to the other is obviously one half of the period of oscillations.
The period of small amplitude oscillations when the particle starts close
to the minimum of the potential $u(z)$, i.e. at $c_0 \to 1$, equals
$2\pi$.  The period monotonically grows to infinity as $c_0 \to 3/2$ or
$0$.  Therefore, if $\Lambda < \pi$ there are no $\xi$-dependent
solutions, and $z_s (\xi) =1$.  At $\Lambda > \pi$ we have an additional
saddle point $z_s (\xi)$ corresponding to a particle moving from $z_s
(0)=c_0$ to $z_s (\Lambda) =c_+$.

Equation~(\ref{equation_1D}) can be solved analytically in terms of the
elliptic integrals.  In particular, the inverse function of $z_s (\xi)$
has the form
\begin{equation}
\label{saddle1D}
\xi (z_s) = \frac{\sqrt{6}}{\sqrt{c_0 - c_-}}
F \left(\arcsin \sqrt{ \frac{c_0 - z_s }{c_0 - c_+} }, 
\sqrt{\frac{c_0 - c_+}{c_0 - c_-}}\, \right),
\end{equation}
where $F(\varphi ,k)$ is the elliptic integral of the first
kind.\cite{Gradshteyn} Using Eq.~(\ref{saddle1D}) the length of the strip
$\Lambda$ can be expressed in terms of $c_0$ as
\begin{equation}
\label{c0}
\Lambda = \frac{\sqrt{6}}{\sqrt{c_0 - c_-}}
F \left(\frac{\pi}{2}, \sqrt{\frac{c_0 - c_+}{c_0 - c_-}}\, \right).
\end{equation}
Solving equation~(\ref{c0}) with respect to $c_0$ one can obtain the
dependence $c_0 (\Lambda)$.  Substituting it into Eq.~(\ref{saddle1D}) and
inverting $\xi (z_s)$ one obtains the saddle point $z_s (\xi)$ for a given
$\Lambda$.  The saddle-point solutions $z_s(\xi)$ for several values of
$\Lambda$ are shown in Fig.~\ref{fig:saddle1D}(a).

For a strip of infinite length the boundary conditions take the form $c_0
= z_s (0) = 3/2$ and $c_+ = z_s (\infty) = 0$.  Then
expression~(\ref{saddle1D}) can be significantly simplified and yields
\begin{equation}
\label{saddle_1Dinf}
z_s (\xi) = \frac{3}{2\cosh^{2} (\xi /2)}.
\end{equation}
This solution can be easily verified by substitution into
Eq.~(\ref{equation_1D}).

The mean switching time $\tau$ is given by
\begin{equation}
\label{tau1exp}
\tau = \tau^{*} e^{F\{z_s\}},
\end{equation} 
where $\tau^{*}$ is a preexponential factor.  Using the expression for
$F\{z\}$, Eq.~(\ref{P0}), we find
\begin{equation}
\label{Fsaddle}
F\{z_s\} = U_1 I(\Lambda),\quad
I(\Lambda)=\! \int^{\Lambda}_0 \! d\xi 
\left(\frac{{z'}_{s}^{2}}{2} + \frac{z_{s}^{2}}{2} 
- \frac{z_{s}^{3}}{3} \right),
\end{equation}
At $\Lambda < \pi$, the only saddle point is $z_s =1$, and therefore
$I(\Lambda) = \Lambda/6$.  At $\Lambda > \pi$ the saddle point
solution $z_s (\xi)$ given by Eq.~(\ref{saddle1D}) corresponds to a
smaller value of $I$.  In particular, at $\Lambda\to\infty$ the
solution~(\ref{saddle_1Dinf}) gives $I = 3/5$.  In the intermediate
region the integral $I(\Lambda)$ can be evaluated analytically,
\begin{eqnarray}
\label{intI}
I &=& \frac{\Lambda}{60}\left[ 4c_+^3 - 6(c_+^2 +c_+ + c_0 - c_- ) + 9 
\right]
\nonumber \\ 
&&+ \frac{\sqrt{6}}{5}\sqrt{c_0 - c_-}\, 
E\left( \sqrt{\frac{c_0 - c_+}{c_0 - c_-}}\,\right).
\end{eqnarray}
Here $E(k)$ is the complete elliptic integral of the second
kind.\cite{Gradshteyn} The dependence $I(\Lambda)$ is plotted in
Fig.~\ref{fig:saddle1D}(b).

Note that apart from the saddle point described by Eq.~(\ref{saddle1D}),
for $\Lambda > 2\pi$ there are additional $\xi$-dependent saddle points.
For example, there is a $\xi$-dependent solution corresponding to a
particle moving from $z_s (0)=c_0$ to $z_s (\Lambda) =c_+$ and back to
$c_0$.  This saddle point is responsible for the processes of switching
inside the strip, see Sec.~\ref{inside1D}.  At $\Lambda > 3\pi$ another
saddle point appears which corresponds to a particle moving from $c_0$ to
$c_+$, returning to $c_0$, and back to $c_+$.  In general, for $\Lambda$
between $\pi m$ and $\pi (m+1)$ there exist $m$ different $\xi$-dependent
solutions.  However, the additional saddle points give larger values of
$F\{z_s\}$ and thus do not affect the switching time.

\section{\label{end} Prefactor of the switching time}

The exponential dependence of the mean switching time~(\ref{tau1exp}) was
obtained from the stationary distribution function $P_0\{n\}$ of electron
density.  However, to calculate the prefactor $\tau^{*}$, understanding of
the time evolution of the distribution function $P\{n(x),t\}$ is also
required.  When electrons tunnel in or out of the well, the density $n(x)$
changes in very small increments.  Thus the dynamics of $P\{n(x),t\}$ is
described by a Fokker-Planck equation.~\cite{dbrts2D} If the strip is very
short, the density in the quantum well is uniform, and the system dynamics
is described by a single variable $n$.  In this case the Fokker-Planck
equation for $P(n,t)$ essentially coincides with the one for a small
two-dimensional sample.  Then by using conventional
techniques~\cite{vanKampen} the prefactor of the switching time can be
found~\cite{dbrts2D} as
\begin{equation}
\label{tau0*}
\tau_0^{*} = \frac{2\pi}{b\sqrt{\alpha\gamma}}.
\end{equation}
This result is correct as long as $\Lambda \ll 1$.  

In longer strips the density fluctuates, and therefore the Fokker-Planck
equation with one variable $n$ cannot adequately describe the evolution of
the distribution function $P$.  In this case $P$ is a functional of
$n(x)$, and the Fokker-Planck equation is multidimensional.  Then to study
the decay of metastable current states one should use a more sophisticated
method.

\subsection {\label{sec:FPE} Fokker-Planck approach to current switching in DBRTS}

To find the prefactor of the mean switching time for the system described
by the multidimensional Fokker-Planck equation, we use the first passage
time technique.~\cite{Matkowsky, Hanggi} This method is applied to the
Fokker-Planck equation in the form
\begin{subequations}
\label{FPE}
\begin{eqnarray}
\label{dPdt}
\!\!\!\!\!\!\!\!\! &&\frac{\partial P({\bf x},t)}{\partial t} 
= {\cal L}\, P({\bf x},t),
\\ 
\label{L}
\!\!\!\!\!\!\!\!\! &&{\cal L} 
= -\sum\limits_{i}\frac{\partial }{\partial x_{i}} K_{i} (\mathbf{x})
+ \sum\limits_{i,j}\frac{\partial^2}{\partial x_{i} \partial x_{j}} 
D_{ij}({\bf x}),
\end{eqnarray}
\end{subequations}
where the matrix $D_{ij}$ represents the generic diffusion coefficient,
and $\mathbf{K}$ is the drift field.  Assuming that the system has a
metastable state, one can consider its domain of attraction $\Omega$ with
the domain boundary $\partial\Omega$ being the separatrix of the field
$\mathbf{K}$.  For the stochastic process described by Eq.~(\ref{FPE}),
the mean time of the first passage out of the domain $\Omega$ has been
found in Refs.~\onlinecite{Matkowsky, Hanggi}.  The mean switching time is
obtained as doubled mean first-passage time and takes the
form,~\cite{Hanggi}
\begin{equation}
\label{Hanggi}
\tau = - \frac{2\int_{\Omega}d^{d}x\, 
P_0 ({\bf x})}{\sum\limits_{i} \int_{\partial \Omega} dS_{i}\,
\sum\limits_{j} D_{ij}({\bf x})P_0 ({\bf x})
\frac{\partial f({\bf x})}{\partial x_j}}.
\end{equation}
Here $P_0$ is the stationary solution of Eq.~(\ref{FPE}).  The form
function $f({\bf x})$ is defined as a stationary solution of the adjoint
equation,
\begin{equation}
\label{f}
{\cal L}^{\dagger}f({\bf x},t) = \sum\limits_{j}\left(K_{j}({\bf x}) 
+ \sum\limits_{i} D_{ij}({\bf x}) \frac{\partial}{\partial x_{i}}\right)
\frac{\partial f({\bf x})}{\partial x_{j}} = 0.
\end{equation}
It satisfies the boundary conditions $f({\bf x}) = 1$ inside the domain
$\Omega$ and $f({\bf x})= 0 $ at the boundary $\partial\Omega$.
 
The Fokker-Planck equation for the case of non-uniform electron density
${n({\bf r})}$ in the well was obtained in Ref.~\onlinecite{dbrts2D}.  It
takes the form
\begin{equation}
\label{FPEgeneral}
\frac{\partial P\{n,t\}}{\partial t}  = \frac{b}{2}
\int\!\! d\mathbf{r}\, \frac{\delta}{\delta n} 
\left( u'(n) - 2\eta\nabla^{2}n 
+ \frac{\delta}{\delta n} \right) P\{n,t\}.
\end{equation}
It is easy to check that the stationary solution of this equation is given
by Eq.~(\ref{P0general}).  

Close to the threshold $u(n)$ can be approximated by the
expansion~(\ref{u}), and the Fokker-Planck equation~(\ref{FPEgeneral}) can
be represented in terms of dimensionless variable $z(\xi)$,
Eq.~(\ref{nz}), as
\begin{equation}
\label{dimensionlessFPE}
\frac{\partial P\{z,\Theta\}}{\partial \Theta} =  
\int\!\! d\xi\, \frac{\delta}{\delta z}
\left( -\frac{d^{2}z}{d\xi^2} + z - z^{2} 
+ \frac{1}{U_1} \frac{\delta}{\delta z} \right) P\{z,\Theta\}.
\end{equation}
where $\Theta = 2\pi t/\tau^{*}_0$ is the dimensionless time.  The
stationary solution $P_0 (z)$ of this equation is $e^{-F}$ with $F\{z\}$
given by Eq.~(\ref{P0}).  Equation~(\ref{dimensionlessFPE}) has an
infinite number of variables, since the density $z(\xi)$ is different at
every point.  From now on we can consider a purely mathematical problem of
decay of a metastable state for the system governed by dimensionless
equation~(\ref{dimensionlessFPE}).  This equation is rather generic and we
expect it to describe other one-dimensional problems of decay of
metastable states.

\subsection {\label{sec:end} Nucleation at an end of a strip}

To find the prefactor of the mean switching time (\ref{tau1exp}) we use
expression~(\ref{Hanggi}).  We evaluate both integrals in
Eq.~(\ref{Hanggi}) in Gaussian approximation.  The integral in the
numerator of Eq.~(\ref{Hanggi}) is dominated by the minimum of $F\{z\}$.
To find the expression for $F$ in the vicinity of its minimum, it is
convenient to use Fourier expansion
\begin{equation}
\label{znearmin}
z(\xi) = \frac{x_0}{\sqrt{\Lambda}} 
+\sum\limits_{i=1}^{\infty} x_{i}\phi_{i}(\xi),\quad
\phi_{i}(\xi) = \sqrt{\frac{2}{\Lambda}} 
\cos\left(\frac{\pi i}{\Lambda} \xi \right).
\end{equation} 
Substituting this expansion into Eq.~(\ref{P0}), we find that up to
quadratic in $x_i$ terms,
\begin{equation}
\label{Fmin}
F(\mathbf{x}) = \frac{U_1}{2} \sum_{i=0}^{\infty} \lambda_i x_i^2, 
\end{equation}
where we defined $\lambda_i \equiv (\pi i /\Lambda)^2 +1$.  Using this
expression for $F$, one can easily evaluate the integral in the numerator
of Eq.~(\ref{Hanggi}),
\begin{equation}
\label{numerator}
\int_{-\infty}^{\infty} \prod_i dx_i \, e^{-F(\mathbf{x})}
= \prod_i \sqrt{\frac{2\pi}{U_1 \lambda_i}}.
\end{equation}

The integral in the denominator of Eq.~(\ref{Hanggi}) is dominated by the
saddle point $z_s(\xi)$ of the functional $F\{z\}$.  To find the
expression for $F$ in the vicinity of the saddle point it is convenient to
expand $z(\xi)$ near $z_s (\xi)$ as
\begin{equation}
\label{znearsaddle}
z(\xi) = z_s (\xi) 
+ \sum\limits_{i=0}^{\infty} {\tilde x}_{i}{\tilde \phi}_{i}(\xi).
\end{equation} 
Here ${\tilde \phi}_{i}(\xi)$ are the normalized solutions of the
eigenvalue problem
\begin{equation}
\label{LAMBDAfinite}
\left(- \frac{d^2}{d \xi^{2}} - 2z_s (\xi) +1 \right){\tilde \phi}_i (\xi) 
= {\tilde \lambda}_i {\tilde \phi}_i (\xi),
\end{equation}
with the boundary conditions
\begin{equation}
\label{BC2}
{\tilde \phi}'_i (0) = 0,\quad {\tilde \phi}'_i (\Lambda) = 0.  
\end{equation}  

Substituting Eq.~(\ref{znearsaddle}) into $F\{z\}$ given by
Eq.~(\ref{P0}), and expanding near $\mathbf{{\tilde x}} = 0$ up to the
second order in ${\tilde x}_i$, we find
\begin{equation}
\label{Psaddle1Dend}
F\{\mathbf{{\tilde x}}\} =
U_1 I(\Lambda) + \frac{U_1}{2}
\sum_{i=0}^{\infty}{\tilde \lambda}_i {\tilde x}_{i}^{2},
\end{equation}
where the first term is given by Eq.~(\ref{Fsaddle}).

Expression~(\ref{Psaddle1Dend}) implies that it is convenient to calculate
the integral in the denominator of Eq.~(\ref{Hanggi}) in terms of
variables ${\tilde x}_i$.  Since the expansion coefficients ${\tilde x}_i$
are related to $x_i$ by orthogonal trasformation, the integrals over $x_i$
in the denominator of Eq.~(\ref{Hanggi}) can be replaced by those over
${\tilde x}_i$.

To calculate the integral in the denominator of Eq.~(\ref{Hanggi}) we also
need to find $D_{ij}$ and $\partial f/\partial {\tilde x}_j$.  They can be
obtained from the $\mathbf{{\tilde x}}$-representation of the
Fokker-Planck equation~(\ref{dimensionlessFPE}).  The adjoint to the
operator ${\cal L}$ of Eq.~(\ref{dimensionlessFPE}) in the
$\mathbf{{\tilde x}}$-representation can be written as
\begin{equation}
\label{Ldagger_1D}
{\cal L}^{\dagger} = \sum_{i=0}^{\infty}
\left( -{\tilde \lambda}_i {\tilde x}_i 
\frac{\partial}{\partial {\tilde x}_i}
+ \frac{1}{U_1}\frac{\partial^2}{\partial {\tilde x}_i^2}\right).
\end{equation}
Here the terms of higher orders in ${\tilde x}_i$ were neglected.

In the denominator of Eq.~(\ref{Hanggi}) the boundary $\partial
\Omega$ is orthogonal to the unstable direction ${\tilde x}_0$ on the
saddle.  Therefore the sum over $i$ reduces to the only term $i=0$.
Using the definition of $D_{ij}$ we find from Eq.~(\ref{Ldagger_1D})
that $D_{ij} = U_1^{-1} \delta_{ij}$.  The sum over $j$ then reduces
to a single term $j=0$, and we need to find only the derivative
$\partial f /\partial {\tilde x}_0$, which is given by
equation~(\ref{f}) in $\mathbf{{\tilde x}}$-representation.  The
latter equation takes the form
\begin{equation}
\label{f_equation}
\left(-{\tilde \lambda}_0 {\tilde x}_0 
+ \frac{1}{U_1}\frac{\partial}{\partial {\tilde x}_0}\right)
\frac{\partial f}{\partial {\tilde x}_0} = 0.
\end{equation}
Since the saddle point is unstable in the ${\tilde x}_0$-direction,
the eigenvalue ${\tilde \lambda}_0$ is negative.  Thus
equation~(\ref{f_equation}) can be solved with the boundary conditions
$f=1$ at ${\tilde x}_0 \to -\infty$ and $f=0$ at the domain boundary
${\tilde x}_0 = 0$ required by the definition of $f(\mathbf{{\tilde
x}})$, cf. Eq.~(\ref{f}).  As a result, we obtain $\partial f
/\partial {\tilde x}_0 = - (2|{\tilde \lambda}_0 |U_1 /\pi)^{1/2}$ at
the saddle point.

Substituting Eqs.~(\ref{numerator}) and~(\ref{Psaddle1Dend}) into
Eq.~(\ref{Hanggi}) we express the switching time in a strip-shaped device
$\tau_s$ as
\begin{equation}
\label{2_1Dend}
\tau_s = \tau^{*}_s e^{U_1 I(\Lambda)},\quad
\tau^{*}_s = \frac{\tau^{*}_0}{\sqrt{|{\tilde \lambda}_0|} }
\sqrt{\frac{{\tilde \lambda}_1}{\lambda_1}} \Upsilon_s,
\end{equation}
where 
\begin{equation}
\label{3_1Dend}
\Upsilon_s = \prod\limits_{i=2}^{\infty}
\sqrt{\frac{{\tilde \lambda}_i}{\lambda_i}}.
\end{equation}
Thus the evaluation of the prefactor of the switching time reduces to
solving the eigenvalue problem~(\ref{LAMBDAfinite}), (\ref{BC2}).

\subsubsection{Very long strip}

We first consider the limit of a very long strip, $\Lambda \gg 1$.  Then
the saddle point $z_s(\xi)$ is given by Eq.~(\ref{saddle_1Dinf}), so the
equation~(\ref{LAMBDAfinite}) takes the form
\begin{equation}
\label{LAMBDAi}
\left(- \frac{d^2}{d \xi^{2}} - \frac{3}{\cosh^2 (\xi /2)} 
+1 \right){\tilde \phi}_i (\xi) 
= {\tilde \lambda}_i {\tilde \phi}_i (\xi),
\end{equation}
The eigenvalue problem~(\ref{LAMBDAi}) with boundary
conditions~(\ref{BC2}) is solved analytically in the
Appendix~\ref{appendix:lambda1D}.  In particular, we find that the
discrete spectrum consists of two bound states with eigenvalues ${\tilde
  \lambda}_0 = -5/4$ and ${\tilde \lambda}_1 =3/4$.  At finite $\Lambda$
the continuous spectrum is discretized according to
\begin{equation}
\label{Lambda_mend}
{\tilde \lambda}_i = 1 +\left(\frac{\pi i}{\Lambda}\right)^{2} \left(1 
- \frac{\delta(\pi i /\Lambda)}{\pi i}\right)^{2}, 
\end{equation}
where the scattering phase shift $\delta (q)$ has the form
\begin{equation}
\label{phaseshift_1D}
\delta (q) 
= \sum\limits_{n=1}^{3} \arctan \frac{n}{2q}. 
\end{equation}

To evaluate $\Upsilon_s$ it is convenient to take the logarithm of
Eq.~(\ref{3_1Dend}), thereby converting the infinite product to a sum.
In the limit $\Lambda\to\infty$, we find
\begin{equation}
\label{Upsilon_e}
\ln \Upsilon_s 
= -\frac{1}{\pi}\int_{0}^{\infty} \frac{q \delta (q)}{1 + q^2} dq 
= \frac12 \ln \frac{2}{15}.  
\end{equation}
Using this result for $\Upsilon_s$ and the fact that in a long strip
$\lambda_1 \to 1$, we find the mean time of switching in a very long strip
\begin{equation}
\label{tau_1Dend}
\tau_e
= \frac{\sqrt{2}}{5}\tau^*_0
\exp\left(\frac{3 U_1}{5} \right).
\end{equation}

It is important to note that the above calculation accounts for only one
of two equivalent saddle points in a strip-shaped device.  Indeed, the
saddle point fluctuation (\ref{saddle_1Dinf}) is positioned near the left
end of the sample, $\xi=0$.  Alternatively, the density fluctuation could
occur at the right end of the device.  This possibility is formally
described by considering the saddle point $z_s(\Lambda-\xi)$.  Since the
two types of processes are equivalent, the total rate of switching at both
ends of the device is $2/\tau_e$.

\subsubsection{\label{sec:finitestrip} Strip of arbitrary length}

Another regime in which analytical expression for $\tau_s$ can be obtained
is that of $\Lambda < \pi$.  In this case the saddle-point density is
uniform, $z_s (\xi) = 1$.  Substituting it into
equation~(\ref{LAMBDAfinite}) we find ${\tilde \lambda}_i = \lambda_i -2$.
Then expression~(\ref{2_1Dend}), (\ref{3_1Dend}) simplifies significantly,
and the prefactor of $\tau_s$ becomes
\begin{equation*}
\tau^{*}_s = \tau_0^{*} \prod_{i=1}^{\infty} 
\sqrt{\frac{(\pi i)^2 - \Lambda^2}{(\pi i)^2 + \Lambda^2} }.
\end{equation*}
Evaluating the infinite product we obtain
\begin{equation}
\label{belowPi}
\tau^{*}_s = \tau_0^{*} \sqrt{\frac{\sin \Lambda}{\sinh \Lambda} }.
\end{equation}
This result generalizes the formula (\ref{Tau0}) for the switching
time in short devices, $\Lambda\ll1$, to the case of any
$\Lambda<\pi$.

\begin{figure}
  \resizebox{.48\textwidth}{!}{\includegraphics{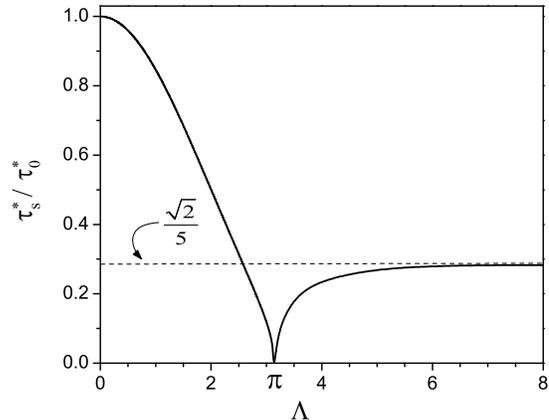}}
\caption{\label{fig:prefactor} Prefactor $\tau^*_s$ of the mean
  switching time in units $\tau^*_0$ vs. the dimensionless length of
  the strip $\Lambda$.  At $\Lambda < \pi$ the prefactor is given by
  Eq.~(\ref{belowPi}), whereas above $\pi$ it is calculated
  numerically.  At $\Lambda\to \infty$ the prefactor approaches
  $\sqrt{2}/5$ in agreement with Eq.~(\ref{tau_1Dend}). }
\end{figure}

For a strip of finite length $\Lambda > \pi$, the saddle point density
$z_s (\xi )$ is given by Eqs.~(\ref{saddle1D}) and (\ref{c0}).  In this
case the eigenvalues ${\tilde \lambda}_i$ are obtained by solving
numerically the eigenvalue problem~(\ref{LAMBDAfinite}) with boundary
conditions~(\ref{BC2}) for a given $\Lambda$.  Substituting them into
expression~(\ref{2_1Dend}), (\ref{3_1Dend}) we find the switching time.
The result for the prefactor of $\tau_s$ as a function of $\Lambda$ is
shown in Fig.~\ref{fig:prefactor}.

Let us discuss the behavior of the prefactor at $\Lambda = \pi$.  At
$\Lambda < \pi$ the functional $F\{z\}$ has only the uniform saddle
point $z_s (\xi) =1$, whereas at $\Lambda > \pi$ there are two saddle
points, the uniform one and the $\xi$-dependent one, see
Fig.~\ref{fig:saddle1D}(a).  These two saddle points merge at $\Lambda
= \pi$.  Formally, this gives rise to the fact that ${\tilde
\lambda}_1$ vanishes at $\Lambda = \pi$, and therefore the prefactor
$\tau^*_s =0$.  We believe that the prefactor remains nonzero at
$\Lambda = \pi$; however, the evaluation of the prefactor in this case
requires a more careful treatment than the Gaussian approximation used
in this paper.  We leave this problem for future work.

It is worth mentioning that the non-monotonic behavior of the
prefactor, see Fig.~\ref{fig:prefactor}, does not result in
non-monotonic dependence of the switching time on sample length
$\Lambda$.  The reason is that the prefactor $\tau^*_s$ is multiplied
by the very large exponential [see
Eqs.~(\ref{tau1exp})-(\ref{Fsaddle}) and Fig.~\ref{fig:saddle1D}(b)]
which monotonically grows with $\Lambda$.  Therefore, the
expression~(\ref{tau1exp}) with the prefactor $\tau^*_s$ and
exponent~(\ref{Fsaddle}) is monotonic everywhere except for the very
narrow region $\pi -3/U_1 \lesssim \Lambda < \pi$.  As discussed in
the previous paragraph, a more accurate evaluation of the prefactor is
required to evaluate the prefactor near $\Lambda=\pi$, and we expect
that it would restore the monotonicity of $\tau$ as the function of
$\Lambda$.

\subsection{\label{inside1D} Nucleation inside long strips and in rings of
  large circumference}

Apart from the ends of a strip, nucleation can occur inside the sample.
Such processes are most important in ring-shaped samples, which have no
ends.  Thus to study interior switching we model the sample by a strip
with periodic boundary conditions.  In the following it will be convenient
to consider a strip of length $2\Lambda$ with boundary conditions
$z(-\Lambda)=z(\Lambda)$, $z'(-\Lambda)=z'(\Lambda)$.

To obtain the switching time we first need to find the saddle point of
$F\{z \}$.  In the mechanical analogy used in Sec.~\ref{sec:1D} [see also
the inset of Fig.~\ref{fig:saddle1D}(a)] the periodic boundary conditions
for the saddle-point equation correspond to a full period of oscillations
of the particle, rather than half period as in the case of nucleation at
the ends in Sec.~\ref{sec:end}.  One such solution is given by $z_s (|\xi
|)$ on the interval $-\Lambda < \xi < \Lambda$, where $z_s (\xi )$ is
defined by Eqs.~(\ref{saddle1D}) and (\ref{c0}), see also
Fig.~\ref{fig:saddle1D}(a).  [Additional solutions are obtained by shifts
$z_s (\xi )\to z_s (\xi +\Delta\xi )$.]  To find the exponent of the
switching time, we need to calculate $F\{z_s \}$ using Eq.~(\ref{P0}).
Since the saddle-point solution $z_s (|\xi |)$ is a symmetric function on
the interval $-\Lambda < \xi < \Lambda$, the integral in $F\{z_s \}$ is
doubled compared to that of end switching.  Thus the time of the switching
inside the strip takes the form
\begin{equation}
\label{Fsaddle_1D}
\tau_r
= \tau^{*}_r e^{2 I(\Lambda) U_1},
\end{equation}
cf. Eq.~(\ref{2_1Dend}).  

The calculation of the prefactor $\tau^{*}_r$ is similar to the one for
the switching at an end of the strip (Sec.~\ref{sec:end}).  As in
Sec.~\ref{sec:end} it is convenient to evaluate the denominator of
Eq.~(\ref{Hanggi}) by expanding $z(\xi)$, Eq.~(\ref{znearsaddle}), where
${\tilde \phi}_{i}(\xi)$ are again the eigenfunctions of
equation~(\ref{LAMBDAi}), but now with periodic boundary conditions
\begin{equation}
\label{periodBC}
{\tilde \phi}_i (-\Lambda) = {\tilde \phi}_i (\Lambda),\quad
{\tilde \phi}'_i (-\Lambda) = {\tilde \phi}'_i (\Lambda).
\end{equation}
The eigenvalue problem~(\ref{LAMBDAi}), (\ref{periodBC}) is solved in the
Appendix~\ref{appendix:lambda1D}.  The discrete spectrum consists of one
negative eigenvalue ${\tilde \lambda}_0 = -5/4$, one zero eigenvalue
${\tilde \lambda}_1 = 0$, and ${\tilde \lambda}_2 = 3/4$; all other
eigenvalues are positive and belong to the quasicontinuous spectrum,
${\tilde \lambda}_i \geq 1$ for $i>2$.  In the denominator of
Eq.~(\ref{Hanggi}) the integration over the amplitudes ${\tilde
  x}_2,{\tilde x}_3,{\tilde x}_4,\dots$ of the modes with positive
eigenvalues is easily performed in Gaussian approximation.  The
integration over the amplitude ${\tilde x}_1$ of the zero mode is less
trivial.

Let us discuss the physical origin of the zero mode.  The functional $F\{z
\}$ is translationally invariant on a ring, and therefore $F\{z_s \}$ does
not change if the saddle-point solution is shifted, $z_s (\xi )\to z_s
(\xi +\Delta\xi )$.  In other words, the deformation $\delta z_s =
\Delta\xi z'_s (\xi)$ does not affect $F$.  On the other hand, according
to Eqs.~(\ref{znearsaddle}) and~(\ref{Psaddle1Dend}) the deformation
$\delta z_s = {\tilde x}_1 {\tilde \phi}_1 (\xi)$ of the saddle-point
solution does not change $F$ either, because ${\tilde \lambda}_1 =0$.
Since the zero mode is unique, we conclude
\begin{equation}
\label{z'phi}
\Delta\xi z'_s (\xi) = {\tilde x}_1 {\tilde \phi}_1 (\xi).
\end{equation}
Then it follows that~\cite{footnote1} $z'_s (\xi) = c_1^{} {\tilde \phi}_1
(\xi)$.  The constant $c_1^{}$ can be found from the normalization
condition for the eigenfunctions ${\tilde \phi}_i (\xi)$,
\begin{equation}
\label{c1}
c_1^{} = \sqrt{\int [z'_{s}(\xi)]^{2}\, d\xi}.
\end{equation}
Using Eq.~(\ref{z'phi}) and the relation $z'_s (\xi) = c_1^{} {\tilde
  \phi}_1 (\xi)$, the integral over the amplitude ${\tilde x}_1$ of the
zero mode takes the form
\begin{equation}
\label{zeromode_1D}
\int\! d {\tilde x}_1^{} = c_1 \int\! d(\Delta\xi ) 
= 2 c_1^{} \Lambda.
\end{equation}

Expressions for $\partial f /\partial {\tilde x}_0$ and $D_{ij}$ were
obtained in Sec.~\ref{end} independently of the exact form of the
saddle-point density and are therefore still applicable.  Substituting
expression~(\ref{numerator}) for the numerator of Eq.~(\ref{Hanggi}) and
using Eq.~(\ref{zeromode_1D}) in the denominator, we find
\begin{equation}
\label{2_1D}
\tau^{*}_r = \frac{\sqrt{\pi}\,\tau^{*}_0}{
\sqrt{2|{\tilde \lambda}_0|\lambda_1} c_1 \Lambda\sqrt{U_1}} 
\sqrt{\frac{{\tilde \lambda}_2}{\lambda_2}} \Upsilon_r .
\end{equation}
Here the infinite product $\Upsilon_r$ is similar to $\Upsilon_s$
evaluated in Sec.~\ref{sec:end}, Eq.~(\ref{3_1Dend}), but with the
eigenvalues $\lambda_i$ and ${\tilde \lambda}_i$ calculated with periodic
boundary conditions~(\ref{periodBC}).  It is shown in the
Appendix~\ref{appendix:lambda1D} that at $\Lambda\gg 1$ the
quasicontinuous spectrum of the eigenvalue problem~(\ref{LAMBDAi}),
(\ref{periodBC}) is still given by Eq.~(\ref{Lambda_mend}), but becomes
doubly degenerate.  Therefore, in the limit of an infinite strip the
product $\Upsilon_r$ can be found as
\begin{equation}
\label{Upsilon}
\Upsilon_r = \Upsilon_s^2 = \frac{2}{15},  
\end{equation}
cf. Eq.~(\ref{Upsilon_e}).  Also, substituting
expression~(\ref{saddle_1Dinf}) for $z_s (\xi)$ into Eq.~(\ref{c1}) we
find $c_1^{} = \sqrt{6/5}$ at $\Lambda\gg 1$.  Upon substitution of these
results into Eq.~(\ref{2_1D}) the mean time of switching inside a long
strip takes the form,
\begin{equation}
\label{tau_1D}
\tau_{i} = \frac{\sqrt{\pi}}{15 \Lambda} \frac{\tau_0^*}{\sqrt{U_1}} 
\exp\left( \frac{6U_1}{5} \right).
\end{equation}
where $U_1$ and $\tau_0^*$ are given respectively by Eqs.~(\ref{U1}) and
(\ref{tau0*}), and $\Lambda = L/2r_0$.  To obtain the exponent we used the
fact that the integral $I=3/5$ at $\Lambda\gg 1$.

In the case of a device of strip geometry, switching can be initiated both
at the ends and inside the sample.  The exponent of $\tau_i$ is a factor
of 2 larger than the exponent of $\tau_e$ [see Eq.~(\ref{tau_1Dend})],
which makes the switching at the ends generally more favorable.  On the
other hand, the rate $1/\tau_i$ is proportional to the strip length $L$.
Therefore in very long strips interior switching denominates.

\subsection{\label{ring} Nucleation in rings of arbitrary circumference}

Expression~(\ref{tau_1D}) was obtained in the limit $L \gg r_0$.  In
this section we discuss the case of a ring of finite circumference $L
\lesssim r_0$.  We will consider separately the cases $L< 2\pi r_0$
and $L> 2\pi r_0$.  We will use the expression~(\ref{Fsaddle_1D}) for
the exponential of the switching time, which was obtained for an
arbitrary circumference of the ring.

Let us start with the case $L< 2\pi r_0$.  In dimensionless units it
corresponds to the problem of a ring with the circumference $2\Lambda
< 2\pi$.  Similarly to the case of a strip of length $\Lambda < \pi$,
see Sec.~\ref{sec:finitestrip}, the saddle-point density is uniform,
$z_s (\xi) = 1$, and from equation~(\ref{LAMBDAfinite}) we obtain
${\tilde \lambda}_i = \lambda_i -2$.  Analogously to the
result~(\ref{2_1Dend}), (\ref{3_1Dend}) for the strip geometry the
prefactor takes the form
\begin{equation}
\tau^{*}_{r} = \frac{\tau_0^{*}}{\sqrt{|{\tilde \lambda}_0|}} 
\prod_{i=1}^{\infty} \sqrt{\frac{{\tilde \lambda}_i}{\lambda_i}}.
\end{equation}
Since the boundary conditions for the ring are periodic,
Eq.~(\ref{LAMBDAfinite}) has two types of solutions, $\cos(\pi i
\xi/\Lambda)$ and $\sin (\pi i \xi/\Lambda)$.  The two solutions have
the same eigenvalues for any $i>0$, and thus all egenvalues except for
${\tilde \lambda}_0$ are doubly degenerate.  Repeating calculations
similar to those for Eq.~(\ref{belowPi}), we find
\begin{equation}
\label{belowPiRing}
\tau^{*}_{i} = \tau_0^{*} \frac{\sin \Lambda}{\sinh \Lambda}.
\end{equation}
This expression is similar to the result~(\ref{belowPi}) for the strip
of length $\Lambda < \pi$.  The absence of the square root in the
right-hand side of Eq.~(\ref{belowPiRing}) is due to the double
degeneracy of the eigenvalues.

\begin{figure}
  \resizebox{.48\textwidth}{!}{\includegraphics{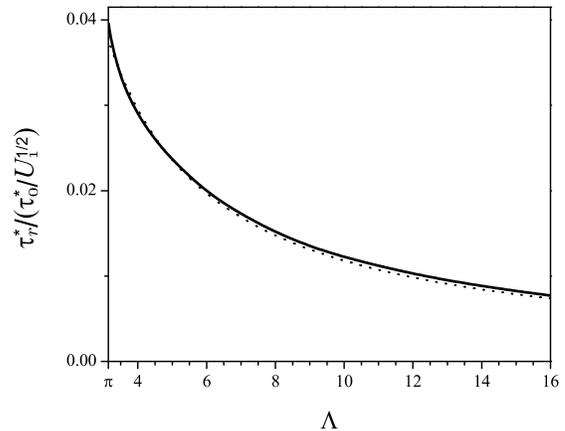}}
\caption{\label{fig:prefactor_ring} Prefactor $\tau^*_r$ of the mean
  switching time in units $\tau^*_0/\sqrt{U_1}$ vs. the dimensionless
  circumference of the ring $2\Lambda$ above $\Lambda = \pi$.  The
  dotted line shows the asymptote $\sqrt{\pi}/15\Lambda$ of $\tau^*_r$
  at $\Lambda\to\infty$, see Eq.~(\ref{tau_1D}).}
\end{figure}

At $\Lambda > \pi$ one needs to solve the eigenvalue
problem~(\ref{LAMBDAfinite}) with nonuniform saddle point $z_s (\xi)$,
Eqs.~(\ref{saddle1D}) and (\ref{c0}), which can be done numerically.
Then one substitutes the eigenvalues ${\tilde \lambda}_i$ into
Eq.~(\ref{2_1D}) to find the prefactor $\tau^{*}_r$.  As we found in
the Appendix, in the limit $\Lambda\to\infty$ the three lowest
eigenvalues ${\tilde \lambda}_i$ are nondegenerate whereas the rest of
them are doubly degenerate, because the potential $-2z_s (\xi)$ is
then reflectionless.  Unexpectedly, our numerical calculation shows
that the same property holds for any $\Lambda > \pi$.  Then similar to
Eq.~(\ref{Upsilon}) we find $\Upsilon_r = \Upsilon_s^2$, where
$\Upsilon_s$ defined by Eq.~(\ref{3_1Dend}) was computed in
Sec.~\ref{sec:finitestrip}.  The prefactor $\tau^{*}_{i}$ above
$\Lambda = \pi$ is plotted as a function of $\Lambda$ in
Fig.~\ref{fig:prefactor_ring}.  One can see in
Fig.~\ref{fig:prefactor_ring} that at $\Lambda\to \infty$ the
prefactor approaches $0$ as $1/\Lambda$ in agreement with
Eq.~(\ref{tau_1D}).

Contrary to the case of a strip, see Fig.~\ref{fig:prefactor}, the
prefactor $\tau_r^{*}$ does not vanish as $\Lambda$ approaches $\pi$
from above.  This is due to the fact that at $\Lambda \to \pi$ small
${\tilde \lambda}_2$ in the numerator of Eq.~(\ref{2_1D}) is
compensated by small $c_1$ in the denominator.  To see this let us
consider the saddle point solution close to $\Lambda = \pi$.  At
$\Lambda = \pi + \epsilon$ with $\epsilon\ll 1$ this saddle point can
be expanded up to its second harmonic as
\begin{equation}
z_s = 1 + \kappa_0 +\kappa_1 \cos (\pi\xi/\Lambda) 
+\kappa_2 \cos (2\pi\xi/\Lambda), 
\end{equation}
where $\kappa_0 = -\kappa_1^2 /2$, $\kappa_2 = \kappa_1^2 /6$, and
$\epsilon = 5\pi\kappa_1^2/12$.  Then from Eq.~(\ref{c1}) we obtain $c_1 =
\sqrt{\pi} \kappa_1$, and calculating ${\tilde \lambda}_2$ up to second
order in $\kappa_1$ we find ${\tilde \lambda}_2 = 5\kappa_1^2/3$.  As a
result the ratio $\sqrt{{\tilde \lambda}_2}/c_1$ in Eq.~(\ref{2_1D})
remains finite, and the prefactor at $\Lambda \to \pi^+$ takes the form
\begin{equation}
\tau^*_r = \frac{\sqrt{5}}{2\sqrt{6} \sinh \pi} 
\frac{\tau^*_0}{\sqrt{U_1}}.
\end{equation}

Similar to the case of a finite strip, the singular behavior of
$\tau^*_r$ near $\Lambda = \pi$ is an artifact of Gaussian
approximation.  We expect that a more careful treatment will show that
the prefactor remains nonzero and continuous at $\Lambda = \pi$.

\section{\label{discussion1D} Discussion}

We have studied the mean time $\tau$ of switching from the metastable to
the stable state in one-dimensional double-barrier resonant tunneling
structures.  We calculated both the exponentials and prefactors of $\tau$
for the strip and ring geometries of the sample.  In this section we
discuss the behavior of the mean switching time in a strip-shaped sample
depending on the distance from the threshold and the structural parameters
of DBRTS.

In the very vicinity of the threshold the exponent of the mean switching
time is of order unity and the metastable state decays very rapidly.  As
the voltage $V$ is tuned further inside the bistable region, the decay
time becomes exponentially long.  Our results are applicable in this
regime.  Thus the exponents in the expressions for the switching time in
the regime of short strip, Eq.~(\ref{Tau0}), and in the regime of long
strip, Eqs.~(\ref{tau_1Dend}) and (\ref{tau_1D}), must be much greater
than unity.  To check whether these conditions are satisfied, it is
convenient to introduce a bias-independent characteristic length
\begin{equation}
\label{d1}
d \equiv r_0 \left( \frac{3U_1}{5} \right)^{1/5} 
= \left(\frac{24}{5} \frac{w\eta^{3}}{\gamma^{2}} \right)^{1/5}
\end{equation}
chosen in such a way that the exponent in Eq.~(\ref{tau_1Dend}) takes
the simple form $(d/r_0 )^5$.
% where we used the definition~(\ref{U1}) of $U_1$,

\begin{figure}
  \resizebox{.48\textwidth}{!}{\includegraphics{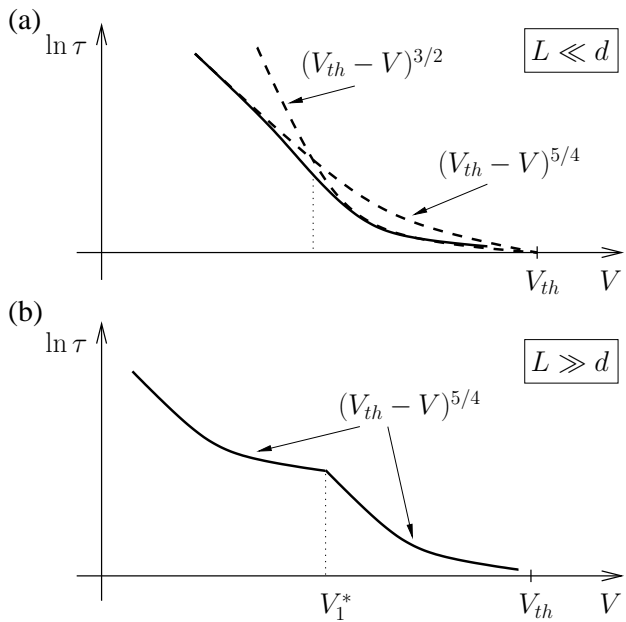}}
\caption{\label{fig:regions1D}(a) Logarithm of the mean switching time
  $\tau$ vs. voltage at $L\ll d$.  Near the threshold there is a region of
  $3/2$-power law dependence of $\ln\tau$ (regime of short strip), then
  follows the region of $5/4$-dependence corresponding to the switching at
  the ends in the regime of long strip. (b) Logarithm of the switching
  time $\tau$ vs. voltage at $L\gg d$.  Close to the threshold one first
  observes $5/4$-power law behavior corresponding to the interior
  switching, then at voltage below $V_1^*$ follows the region of
  $5/4$-dependence corresponding to switching at the ends.}
\end{figure}

The exponent of the mean switching time in the short-strip regime
[Eq.~(\ref{Tau0})] can be expressed in terms of $d$ and $r_0$ as
\begin{equation}
\label{exp_small1D}
\frac{4Lw\alpha^{3/2}}{3\gamma^{1/2}} 
= \frac{5}{18} \frac{L}{r_{0}}\left( \frac{d}{r_0}\right)^{5}.
\end{equation}
This exponent is much greater than unity at $r_0 \ll (L d^5)^{1/6}$.
Since the regime of short strip is defined by $r_0\gg L$, it is present
only if $L\ll d$.  In this case, starting at voltage difference $V_{th} -
V$ corresponding to $r_0\sim (Ld^5)^{1/6}$, there is a region of
$3/2$-power law dependence of $\ln\tau$ corresponding to the short-strip
regime.  Then, as $V_{th} - V$ reaches the value corresponding to $r_0
\sim L$, follows the region of $5/4$-dependence of $\ln\tau$ for the
long-strip regime, see Fig.~\ref{fig:regions1D}(a).

In the latter case the switching can be initiated either inside or at each
of the two ends of the sample.  Summing the rates of these processes we
find
\begin{equation}
\frac{1}{\tau} = \frac{2}{\tau_e} + \frac{1}{\tau_i}.
\end{equation}
To determine the type of $5/4$-dependence that can be observed in the
regime of long strip, $r_0 \ll L$, one should compare the rates of
switching inside and at the ends of the strip.  Using
Eqs.~(\ref{tau_1Dend}) and (\ref{tau_1D}), the ratio of the rates
$\tau_{e}^{-1}$ and $\tau_{i}^{-1}$ can be expressed as
\begin{equation}
\label{Re/Ri_1D}
\frac{\tau_{e}^{-1}}{\tau_{i}^{-1}} \sim
\frac{d}{L}
\exp\left[ \left( \frac{d}{r_0} \right)^{5} 
-\frac{7}{2} \ln\frac{d}{r_0} \right].
\end{equation}
Using this expression we conclude that at $L\ll d$ the switching always
initiates at the ends rather than inside the strip.  Thus the region of
$5/4$-dependence is described by Eq.~(\ref{tau_1Dend}).

At $L\gg d$ there is no region corresponding to the regime of a short
strip.  One can only observe $5/4$-power law dependence of $\ln\tau$
corresponding to the long-strip regime.  Analyzing Eq.~(\ref{Re/Ri_1D})
one can see that at $L\gg d$ there are two distinct regions of $5/4$-power
law dependences.  At $r_0 \lesssim d$ and very large $L$ the switching
initiates inside the strip.  At very small $r_0$ the exponential in
Eq.~(\ref{Re/Ri_1D}) becomes very large, and therefore the switching takes
place at the ends.  The voltage $V_{1}^{*}$, at which the crossover
between these two regions occurs, is given by the condition $\tau_{i}^{-1}
= \tau_{e}^{-1}$ applied to Eq.~(\ref{Re/Ri_1D}),
\begin{equation}
V_1^{*}\approx V_{th} - \frac{\eta^2}{\gamma a d^4} 
\left(\ln\frac{L}{d} \right)^{4/5}.
\end{equation}

Thus we conclude that in the case of $L\gg d$, starting at $V$
corresponding to $r_0\sim d$, one first observes $5/4$-power law
dependence of $\ln\tau$ on voltage corresponding to the switching
inside the strip, then at $V$ below $V_1^*$ follows the region of
$5/4$-dependence corresponding to the switching at the ends, see
Fig.~\ref{fig:regions1D}(b).

\begin{acknowledgments} 
  The authors are grateful to S.~W. Teitsworth and M.~B.  Voloshin for
  fruitful discussions.  O.A.T. acknowledges the hospitality of
  Argonne National Laboratory where part of this work was performed.
  This work was supported by NSF Grant No. DMR-0214149 and the
  U.S. DOE, Office of Science, under Contract No. W-31-109-ENG-38.
\end{acknowledgments}

\appendix*

\section{\label{appendix:lambda1D} Eigenfunctions and eigenvalues of equation~(\ref{LAMBDAi})}

In this Appendix we find the eigenfunctions of discrete spectrum and the
phase shifts of continuous spectrum of equation~(\ref{LAMBDAi}), as well
as the respective eigenvalues ${\tilde \lambda}_i$.  In order to solve
equation~(\ref{LAMBDAi}) one needs to diagonalize the Hamiltonian
\begin{equation}
\label{epsiloni} 
H_m = -\frac{d^{2}}{d \xi^2} - \frac{m(m+1)}{4\cosh^2 (\xi/2)}, 
\end{equation}
with $m=3$.
  
The eigenstates of the Hamiltonian~(\ref{epsiloni}) can be obtained using
algebraic technique called supersymmetric quantum mechanics, see
Refs.~\onlinecite{ShifmanVoloshin, Witten, SchwablBook}.  We begin by
introducing the raising and lowering operators
\begin{equation}
\label{operators}
a^{\pm} = \mp \frac{d}{d \xi} + \frac{m}{2}\tanh \frac{\xi}{2}.  
\end{equation}
Hamiltonians $H_m$ and $H_{m-1}$ can be expressed in terms of $a^{+}$ and
$a^{-}$ as
\begin{subequations}
\label{hamiltonians}
\begin{eqnarray}
\label{aa+}
H_m = a^{+}a^{-} -\frac{m^2}{4},\\
\label{a+a}
H_{m-1} = a^{-}a^{+} - \frac{m^2}{4}.
\end{eqnarray}
\end{subequations}

Let us consider the function $\Psi(\xi) = 1/\cosh^m (\xi/2)$.  It
satisfies the condition $a^{-}\Psi (\xi) =0$, and therefore $\Psi$ is an
eigenfunction of the Hamiltonian $H_m$ with eigenvalue $\epsilon =
-m^2/4$.  Since $\Psi(\xi)$ has no zeros, it describes the ground state of
$H_m$.  Thus we find the ground-state wave function and energy
\begin{equation}
\label{groundPsi}
\Psi_0^{(m)}(\xi) = \sqrt{\frac{\Gamma \left( m+\frac{1}{2} \right)}{2
\sqrt{\pi}\,\Gamma (m)}} \frac{1}{\cosh^m (\xi/2)},\quad 
\epsilon_0^{(m)} = -\frac{m^2}{4}. 
\end{equation}

Applying the raising operator $a^{+}$ to $H_{m-1}\Psi_i^{(m-1)} =
\epsilon_i^{(m-1)} \Psi_i^{(m-1)}$ we find that $a^{+}\Psi_i^{(m-1)}$ are
eigenfunctions of Hamiltonian $H_m$ with eigenvalues $\epsilon_i^{(m-1)}$.
Upon appropriate normalization we obtain
\begin{equation}
\label{normalizedPsi}
\Psi_i^{(m)}(\xi) = \frac{1}{\sqrt{\epsilon_{i-1}^{(m-1)} +m^2 /4}}\, 
a^{+} \Psi_{i-1}^{(m-1)},\quad  
\epsilon_i^{(m)} = \epsilon_{i-1}^{(m-1)}.  
\end{equation}  
Conversely, by acting with $a^{-}$ on $H_m\Psi_i^{(m)} = \epsilon_i^{(m)}
\Psi_i^{(m)}$ with $i > 0$, one finds that $a^{-}\Psi_i^{(m)}$ are
eigenfunctions of Hamiltonian $H_{m-1}$ with eigenvalues
$\epsilon_i^{(m)}$.  Therefore, we conclude that in addition to its
ground-state eigenvalue $\epsilon_0^{(m)}$ given by~(\ref{groundPsi}), the
spectrum of $H_m$ consists of all the eigenvalues of $H_{m-1}$.

Using Eqs.~(\ref{groundPsi}) and~(\ref{normalizedPsi}) one can derive the
eigenfunctions and eigenvalues of $H_m$ from those of $H_{m-1}$ and vice
versa.  To illustrate this, let us consider the Hamiltonian $H_0 =
-(d/d\xi)^2$.  Its eigenfunctions are $e^{\pm iq\xi}$, and the spectrum is
given by $\epsilon_q^{(0)} = q^2$, where $q$ is an arbitrary real wave
number.  Then using Eq.~(\ref{normalizedPsi}) we can find the eigenstates
of Hamiltonian $H_1$.  The continuous spectrum is obviously
$\epsilon_q^{(0)}$, and its normalized eigenfunctions are
\begin{equation} 
\label{f1}
\Psi_q^{(1)}(\xi) = \frac{1}{\sqrt{2\pi (q^2 +1/4)}}  
\left(-\frac{d}{d \xi} +\frac12 \tanh \frac{\xi}{2} \right) 
e^{iq\xi}.
\end{equation}
The ground-state wave function and eigenvalue of $H_1$ are given by
Eq.~(\ref{groundPsi}) with $m = 1$.

Applying the same technique two more times we find the bound states of
$H_3$,
\begin{eqnarray}
\label{f3}
&&\!\!\!\!\!\!\!\!\!\!\!\!\!\!\!\!\!\!\Psi_0^{(3)} (\xi) 
= \frac{\sqrt{15}}{4\sqrt{2}} \frac{1}{\cosh^{3} (\xi/2)},\quad 
\epsilon_0^{(3)} = -\frac94,\\
&&\!\!\!\!\!\!\!\!\!\!\!\!\!\!\!\!\!\!\Psi_1^{(3)} (\xi) 
= \frac{\sqrt{15}}{2\sqrt{2}} \frac{\sinh (\xi/2)}{\cosh^3 (\xi/2)},\quad  
\epsilon_1^{(3)} = -1,\\
\label{f32}
&&\!\!\!\!\!\!\!\!\!\!\!\!\!\!\!\!\!\!\Psi_2^{(3)} (\xi) 
= \frac{\sqrt{3}}{\sqrt{2}\cosh \frac{\xi}{2}} - \frac{5\sqrt{3}}{4\sqrt{2}
\cosh^3 \frac{\xi}{2}},\quad 
\epsilon_2^{(3)} = -\frac14.
\end{eqnarray}
The eigenfunctions of continuous spectrum of $H_3$ are given by
\begin{widetext}
\begin{equation}
\label{PsiCont}
\Psi^{(3)}_q (\xi) = \frac{1}{\sqrt{2\pi(q^2 + 9/4)(q^2 + 1)(q^2 + 1/4)}} 
\left(-\frac{d}{d \xi} +\frac32 \tanh \frac{\xi}{2} \right) 
\left(-\frac{d}{d \xi} +\tanh \frac{\xi}{2} \right) 
\left(-\frac{d}{d \xi} +\frac12 \tanh \frac{\xi}{2} \right) e^{iq\xi}.
\end{equation}
\end{widetext}

Noting that $H_3$ coincides up to a constant with the Hamiltonian of
equation~(\ref{LAMBDAi}), we conclude that ${\tilde \phi}_i (\xi) =
\Psi_i^{(3)} (\xi)$.  Therefore the discrete spectrum of
Eq.~(\ref{LAMBDAi}) has three eigenvalues,
\begin{equation}
\label{discrete_lambda} 
{\tilde \lambda}_0 = -\frac{5}{4},\quad {\tilde \lambda}_1 =0,\quad 
{\tilde \lambda}_2 = \frac{3}{4}.
\end{equation}

The continuous spectrum of Eq.~(\ref{LAMBDAi}) is ${\tilde \lambda}_q = 1
+ q^2$.  The asymptotics of its eigenfunctions~(\ref{PsiCont}) can be
expressed in terms of the scattering phase shifts $\delta (q)$ in the
following way
\begin{equation}
\label{asympt}
{\tilde \phi}_q (\xi) \to \frac{i \operatorname{sgn} q}{\sqrt{2\pi}} 
e^{i q \xi}e^{\pm i \delta (q)},\quad 
\xi\to \pm\infty. 
\end{equation}
Here the absence of the reflected wave illustrates the fact that the
Hamiltonian $H_m$ describes scattering in reflectionless
potential.\cite{LandauQM} Comparing Eqs.~(\ref{PsiCont})
and~(\ref{asympt}) one can see that the phase shifts are given by
Eq.~(\ref{phaseshift_1D}).

For the evaluation of the prefactor of the switching time we need to study
a finite system.  In Sec.~\ref{inside1D} we consider the system of length
$2\Lambda$ with periodic boundary conditions~(\ref{periodBC}).  At
$\Lambda\gg 1$ the eigenfunctions of discrete spectrum of
Eq.~(\ref{LAMBDAi}) with the boundary conditions~(\ref{periodBC}) are
given by Eq.~(\ref{discrete_lambda}).  The quasicontinuous spectrum of a
large finite system can be obtained by applying periodic boundary
conditions~(\ref{periodBC}) to the eigenfunctions ${\tilde \phi}_q$ in the
asymptotic form~(\ref{asympt}).  As a result we find the wave number
quantization
\begin{equation}
\label{q_iteration}
q_l = \frac{\pi l - \delta (q_l )}{\Lambda}, 
\end{equation}
where $l$ is an integer.  Because the phase shift~(\ref{phaseshift_1D}) is
an odd function of $q$, it follows from Eq.~(\ref{q_iteration}) that
$q_{-l} = -q_l$.  Therefore, the eigenfunctions with wave numbers $q_l$
and $q_{-l}$ have the same eigenvalue ${\tilde \lambda} = 1 +q_l^2$.  Thus
the quasicontinuous spectrum of Eq.~(\ref{LAMBDAi}) with periodic boundary
conditions~(\ref{periodBC}) is doubly degenerate.  The two real solutions
with the same ${\tilde \lambda}_q$ can be represented as even and odd
combinations of ${\tilde \phi}_q$ and ${\tilde \phi}_{-q}$.

Transcendental equation~(\ref{q_iteration}) can be solved using iteration
procedure.  At $\Lambda\gg 1$ it is sufficient to perform only the first
iteration.  Substituting the result into ${\tilde \lambda}_i = 1+q_i^2$ we
obtain Eq.~(\ref{Lambda_mend}).

In Sec.~\ref{sec:end} we consider the system of length $\Lambda$ with the
boundary conditions~(\ref{BC2}).  All solutions of this problem can be
obtained from the solutions of the problem with the boundary
conditions~(\ref{periodBC}).  Indeed, any even solution of the eigenvalue
problem~(\ref{LAMBDAi}) with the boundary conditions~(\ref{periodBC})
satisfies the conditions~(\ref{BC2}).  Therefore, from
Eqs.~(\ref{f3})--(\ref{f32}) and~(\ref{discrete_lambda}) one concludes
that the discrete spectrum consists of two eigenvalues $-5/4$ and $3/4$.
Because only even combinations ${\tilde \phi}_q (\xi) +{\tilde \phi}_q
(-\xi) \equiv {\tilde \phi}_q (\xi) -{\tilde \phi}_{-q}(\xi)$ satisfy the
conditions~(\ref{BC2}), the quasicontinuous spectrum is nondegenerate and
given by Eq.~(\ref{Lambda_mend}).


\begin{thebibliography}{99}
\bibitem{footnote0} In this paper we study the problem with the cubic
potential given by Eq.~(\ref{u}).  A similar problem was considered in
Ref.~\onlinecite{ButtikerLandauer81}.  A one-dimensional decay problem
with a quartic potential was addressed in Refs.~\onlinecite{Stein01}
and \onlinecite{Stein05}.
\bibitem{footnote1} By differentiating Eq.~(\ref{equation_1D}), one
  can check explicitly that $z'_s (\xi)$ is indeed a solution of
  equation~(\ref{LAMBDAi}) with ${\tilde \lambda} =0$.
\end{thebibliography}
\end{document}